\documentclass[final,3p,times,sort&compress,twocolumn]{elsarticle}

\usepackage{graphicx}

\usepackage[usenames,dvipsnames]{xcolor}  
\usepackage{amssymb,amsfonts,amsmath}

\usepackage{relsize}
\usepackage{placeins}
\usepackage{capt-of}
\usepackage{hyperref}

\include{strut}

\def\eps{\varepsilon}

\begin{document} 

\title {Thermodynamic description of worldwide distribution \\
  of energy and carbon emission}

\author[LPT]{Klaus M. Frahm}

\author[LPT]{Dima L.~Shepelyansky}

\address[LPT]{\mbox{Unive Toulouse, CNRS, Laboratoire de Physique Th\'eorique, 
Toulouse, France}}

\ead[url]{http://www.quantware.ups-tlse.fr/dima}




\begin{abstract}
Based on  public data, we  analyze the
distributions of energy and carbon emission over world countries
on a scale of the last 40-50 years
using their presentation via Lorenz and Pareto curves.
These curves in rescaled format remain
remarkably stable on this time period
being characterized by high values
of the Gini coefficient indicating 
a strong inequality of energy distribution.
To explain these distributions, 
we introduce the ENergy Thermalization Hypothesis (ENTH)
according to which these distributions result from
the Rayleigh-Jeans (RJ) thermalization and condensation
of agents representing different countries.
We show that this hypothesis provides
an excellent description of 
Lorenz and Pareto curves obtained from
data on the above time period.
It also gives natural grounds for inequality
relating it to the RJ condensation at low energy states.
We additionally trace parallels
with the wealth inequality in the world.
\end{abstract}

\maketitle

\onecolumn


\section{Introduction}
\label{sec1}

The sharing of energy production and consumption, as well as
carbon emission, between world countries is now under
attentive analysis of multiple international institutions
(see e.g. \cite{inst1,inst2,inst3}).  In many cases, 
these institutions provide a public access to the related datasets
for several years for each country (see e.g. \cite{data1,data2,data3}).
In view of the importance of the worldwide energy distribution, it is
highly desirable to understand if it has certain universal features
that may be useful to analyze sustainable multi-energy systems \cite{chertkov}.
In this work, we introduce
the Energy Thermalization Hypothesis (ENTH)
according to which the energy distribution over countries is given
by the thermodynamic law for classical fields
known in physics as the Rayleigh-Jeans (RJ) thermal distribution
\cite{landau,zakharovbook}. In the framework of this concept the 
world countries are 
considered as independent players with multiple
links and nonlinear interactions between each other
that conserve two integrals of motion being
the total system energy and probability norm.
These interactions between players lead to thermalization with
energy and norm distribution between countries
described by the RJ thermodynamic law.

Recently, a similar concept has been proposed as the Wealth
Thermatization Hypothesis (WTH) and it was shown that
the RJ thermalization provides a good description
of wealth distribution of households in countries and the world,
Gross Domestic Product (GDP) of countries, market capitalization of
companies at stock exchange of Hong Kong, Shanghai, London,
bitcoin transactions, world trade between countries \cite{wth1,wth2}.

It is well know that there is a huge wealth inequality in the world
and in the majority of countries
when for the whole world 50\% of the population owns only 2\% of total wealth,
while 10\% (1\%) of the richest population owns 75\% (38\%) of total wealth 
(see e.g. \cite{piketty1,piketty2,boston}). It has been 
shown in \cite{wth1,wth2} that this inequality is naturally
explained by the RJ thermalization
leading to RJ condensation and the formation
of a huge poverty phase of low wealth and a tiny oligarchic phase
that captures a main part of total society wealth.

This RJ condensation phenomenon has been observed for various 
physical systems as e.g.  self cleaning in multimode optical fibers
\cite{wabnitz,picozzi1,chrisrep,picozzi2,babin,ourfiber}.
The RJ condensation is also a specific example
of constraint driven condensation phenomenon known
in statistical mechanics \cite{trizac,satya,marsili}. 
This constraint driven condensation is universal and
exists for such systems  as
coalescence in granular media, jamming in traffic, 
gelation in networks \cite{satya} 
and financial data analysis \cite{marsili}.

In this work, we show that RJ thermalization and condensation 
also describe the distribution of energy and carbon emission over
world countries. To verify the ENTH validity, we compare its predictions
with the Lorenz and Pareto curves taken from public 
datasets \cite{data1,data2,data3} 
for energy and carbon emission distributions over countries on a scale of the 
last 50 years. In economy, the description of wealth distribution over 
households is usually analyzed via the Lorenz curve \cite{lorenz,boston}
which gives the dependence of cumulated 
normalized wealth $0\leq w \leq 1$ on the cumulated normalized fraction of
population or households $0 \leq h \leq 1$.
The case of perfect equipartition of wealth corresponds to
the diagonal $w=h$ and the doubled area between diagonal
and the Lorenz curve $w(h)$ gives the Gini coefficient
$0 \leq G \leq 1$  \cite{gini,boston}. For example,
for the whole world $G=0.842$ in 2021 \cite{piketty2}. 
The Gini coefficient being close to unity corresponds to
a high inequality in a country with a large fraction of poor population
and a tiny oligarchic population fraction that owns
a huge fraction of total wealth \cite{piketty1,piketty2}.
The Pareto curve \cite{pareto} 
stresses more the properties of high revenues.
For energy and carbon emission distributions we
obtain similar high values of the Gini coefficient $G \approx 0.87$-$0.895$ 
corresponding to a strong RJ condensation with a
large phase of energy poor countries and tiny phase of 
energy rich countries. 

In this work, we apply the ENTH approach \cite{wth1,wth2} 
to the description of distribution of energy and carbon emission
which play a role similar to wealth. We show that the RJ thermalization 
and condensation gives a good description of real data on a scale of 
40-50 years.

The paper is constructed as follows:
in Section 2 we describe the construction procedure of Lorenz and Pareto curves
as well as features of the RJ distribution, the RJ theory results for 
energy and carbon, or CO$_2$,  emission distributions over countries are 
presented in Section 3, the discussion of results is given in Section 4 and 
additional figures are given in Appendix. 

\section{Construction of Lorenz and Pareto curves and ENTH theory}

The thermodynamic approach to distribution of money, in human society
was first proposed and studied by
Yakovenko et al. in  \cite{yakovenko1,yakovenko2} 
arguing that distributions of money, income and wealth
are described by the Boltzmann-Gibbs (BG) distribution at a 
certain temperature $T$. 
However, a comparison with real Lorenz datasets shows that
for the BG distribution the Lorenz curve does not depend on temperature and has
a constant Gini coefficient $G=0.5$ that does not correspond to real situations 
where $G$ changes significantly with time and also many
countries like US having a rather high value $G \approx 0.8$ 
(see also \cite{wth2}).
Thus in \cite{yakovenko1,yakovenko2} the BG distribution was modified by hand 
replacing its exponential decay at high revenues 
by a Pareto algebraic tail allowing to
obtain a variation of $G$ that depends on the Pareto curve
parameters. Another feature of the BG description 
is that as for an atomic gas it gives a distribution 
of velocity or energy fluctuations of atoms 
while the average energy of each atom is always equal to
$3T/2$ for all atoms (here the Boltzmann constant is equal to unity). 

In later studies Yakovenko et al. \cite{yakovenko3,yakovenko4}, 
also constructed Lorenz curves for energy and carbon emission
for world countries, but in these works
they considered for each country
energy and carbon emission values per capita
(total value divided by country population). 
We argue that the consideration of country energy
per capita does not correctly describe the 
thermalization process induced by
interactions between countries. 
Indeed, only countries can be considered
as individual players competing for energy
distribution worldwide. In contrast
an individual person of a given country
has practically no influence on this competition since
only collective interactions on a country level affect the energy
distribution in the world. 
For example, for the GDP distribution over countries
it was shown in \cite{wth2} that the Lorenz curves
for GDP per capita have at the top positions
such small countries as Monaco and Liechtenstein
which do not have a significant
influence on the global world economy.
Therefore, we conclude that Lorenz and Pareto curves
should be constructed for total energy consumption 
(or CO$_2$ emission) over countries 
and not for countries per capita. 
This conclusion is confirmed by the results presented below.

Another important features of wealth distributions
were outlined in \cite{boghosian1,boghosian2}
where it was stressed that there are two integrals of
the system evolution 
being the total wealth and total probability norm
related to a number of interacting agents but 
in \cite{boghosian1,boghosian2} the system evolution
was described on the basis of kinetic equations
without any thermalization arguments. 
Also in \cite{boghosian1,boghosian2} it was argued
that it is more correct to analyze the properties of
wealth, instead of money or income, since its variations
satisfy the small-transaction approximation. 
The conservation of two integrals is a rather natural assumption
since the total wealth and population of a country or of the whole
world are only weakly changed on a typical time scale of one year.

In the framework of the ENTH approach, similarly as in \cite{wth1,wth2},
we assume that countries are represented by
a certain number $N$ of interacting agents
($0 \leq m < N$) that are similar to a system of nonlinear 
coupled oscillators. These agents, or oscillators,
have energies levels $E_m$, or frequencies, of linear modes
located in a system energy band $B$: $0 \leq E_m \leq B$. 
The probability to find an agent in a mode $m$ is $\rho_m$ 
defined as a certain long time average of the squared oscillator 
amplitude of the mode $m$ (see \cite{wth1,wth2} for more details). 
Nonlinear interactions between modes lead
to dynamical thermalization as it was demonstrated
for various physical models \cite{wth2,ourfiber,rmtprl,stratif}. 
Due to the conservation of two integrals of motion
the RJ thermalized value of $\rho_m$ (see e.g. \cite{landau,zakharovbook} 
and also \cite{rmtprl} for more specific details) is given by
\begin{equation}
\rho_m = \frac{T}{E_m-\mu} \;\;\; ({\rm RJ}) .
\label{eqrj}
\end{equation}
Here the parameters $T$ and  $\mu(T)$
are the system temperature and
its  chemical potential dependent on $T$.
They are determined by (\ref{eqrj}) and 
the two implicit equations $\sum_m \rho_m =1$ (norm) and 
$\sum_m E_m \rho_m =E$ 
(system energy) related to the two integrals of motion. 
For a given oscillator spectrum 
$E_m$ and a given value of $E\in[E_0,E_{N-1}]$, one can compute numerically 
the physically valid solution of $\mu$ (such that $\rho_m>0$ for all modes) 
using the identity 
$T=(E-\mu)/N$ (see e.g. \cite{wth1,wth2,rmtprl} for some details on this). 
This solution satisfies either $\mu<E_0$ with $T>0$ or $\mu>E_{N-1}$ with 
$T<0$ (here in this work the used values of $E$ to fit real data correspond 
to the regime of $T>0$). 

For small energies $E\sim E_1-E_0$ and related low temperatures 
(and very small values of $|\mu-E_0|$) a macroscopic part 
of the total probability norm 
is condensated at the lowest energy modes $E_k$ (for modest values 
of $N$ condensation on the lowest mode $E_0$ with $\rho_0\gg \rho_m$ 
for $m>0$ is also possible). 
As we will see below, this condensation naturally explains 
the appearance of strong wealth inequality in the associated 
Lorenz curves 
with a huge fraction of poor population and a tiny oligarchic fraction
that captures the main part of total wealth of the world. 
Details of this RJ condensation can be found in \cite{wth1,wth2,stratif} 
and we expect that a similar condensation takes place for 
the energy consumption and carbon emission distributions over countries.

We mention that the RJ distribution (\ref{eqrj}) also 
directly follows from the quantum Bose-Einstein 
(BE) distribution \cite{landau} 
\begin{equation}
 \rho_m=\frac{1}{\exp[(E_m-\mu)/T]-1} \; ({\rm BE}) .
\label{eqbe}
\end{equation}
as a limiting case for high temperature $T \gg  (E_m-\mu)$ but for the 
classical nonlinear oscillator system (see e.g. \cite{wth1,wth2,rmtprl}) 
it is valid for all temperatures. 

For a given spectrum $E_m$ with related RJ probabilities $\rho_m$ 
(\ref{eqrj}), we can construct the Lorenz curve 
by computing the cumulated normalized 
fraction of households as $h(m) =\sum^{m-1}_{k=0} \rho_k$
and the cumulated normalized wealth fraction 
$w(m) = \sum^{m-1}_{k=0} w_k \rho_k/w_s$
where $w_k=E_k$ is the individual wealth/energy consumption (or 
CO$_2$ emission) 
associated to a given agent $k$. Here $w_s$ is the average wealth given by 
$w_s = \sum w_m\rho_m$ which corresponds to the energy parameter $E$ in the 
RJ approach. This procedure provides the Lorenz curve as $N+1$ data points 
$(h(m),w(m))$ for $0\le m\le N$ such that $h(0)=w(0)=0$ and $h(N)=w(N)=1$ 
and it is invariant with respect to a simple energy 
rescaling $E_m\to\alpha E_m$, $E\to \alpha E$. Therefore, the relevant 
parameter is the rescaled energy $\varepsilon = w_s/B=E/B$ where 
$B=E_{N-1}$ is the energy band width of the spectrum (in this work, 
we also assume $E_0=0$). 

For the cases of real data, e.g. of energy consumption 
or carbon emission, 
the Lorenz curve is constructed in a similar way but with uniform 
$\rho_m =1/N_c$ for each country and $w_m$ taken from the databases
\cite{data1,data2,data3}. Here $N_c$ is the number of world countries
(including Greenland) for a given year, usually $N_c \approx 200$.

To  compare the RJ theory (\ref{eqrj}) with the Lorenz curve
from real data, we assume that each country is represented by a certain number
of thermalized agents $N$ distributed over
energy levels $E_k$ being in the energy band $B$.
The simplest assumption is that the density of energy states
$\nu(k) =d k/d E_k$ is constant so that
$0 \leq E_k=k/N < B=1$ for $k=0,1,2,\ldots,(N-1)$. This model
is called the RJ standard (RJS) model \cite{ourfiber,wth1,wth2}. 
For sufficiently small values of $\eps$ it shows RJ condensation 
where a significant fraction of probability is concentrated on a 
small number of modes at minimal $k$ values and it provides Lorenz curves 
that agree on a qualitative level with real data Lorenz curves. 

However, to obtain a more precise description, we use a more refined 
model called the RJ extended (RJE) model with 
$E_k =(e^{ak/N}-1)/(e^a-1)$ where $a$ is 
a real parameter. The limit $a\to 0$ reproduces the RJS model 
with $E_k=k/N$ and for $a>0$ the density of states is decreasing at 
high energies according to 
with $\nu(E_k)=dk/dE_k = N(e^a-1)/[a(1+(e^a-1)E_k]$. Here the global 
energy scale is chosen such that $B=E_{N-1}\approx 1$. 
The decay of $\nu$ at high energies is rather natural since
there are less rich people (or countries) at high wealth/energy values 
and in this work, we mainly present results for the RJE model. 
To recover the Lorenz curve from the RJ theory (\ref{eqrj}) 
we apply the above procedure for at least $N=10000$  agents with 
probabilities $\rho_k$ given by 
(\ref{eqrj}), but in some cases we use even higher $N$ values.
 Then multiple agents describe
real values $\rho_m, w_m$ for a given country.
It is also possible, to work out explicit analytical formula for 
the continuous limit $N\to\infty$ in terms of the chemical potential 
$\mu$ which is determined by an implicit equation as a function of 
$\eps$ and $a$ and the Lorenz curves are very stable for sufficiently 
large values of $N$ (and agree with the continuous limit $N\to\infty$)
at given parameters $a$ and $\eps$ (see details in 
\cite{wth2}).

To compare the RJS and RJE models with real Lorenz curves, we choose 
the parameter $\eps$ such that the RJ and real Lorenz curve have the 
same Gini coefficient $G$. For the RJE model, we also minimize the geometrical 
curve distance between both curves to determine the optimal 
choice for the parameter $a$ 
(under the constraint that for each value of $a$ the parameter $\eps$ 
is recomputed to provide the same Gini coefficient $G$ as the real data). 
This procedure is the same as the one used in \cite{wth1,wth2}. 

The Lorenz curve highlights the wealth or energies of
the poor to medium wealth population fraction/countries. 
Another quantity, which focuses more on the details for the very 
rich population fraction is the cumulative distribution function (CDF) 
$C(w_m)$ which gives the fraction of households (companies, people, 
countries etc.) having a wealth larger than $w_m$. This quantity is 
typically used in economy (see e.g. \cite{yakovenko1,yakovenko2}) 
and it can be directly obtained 
by drawing $C(w_m)=1-h(m)$ versus $w_m$ (for both cases 
of real data and RJ models). For this, typically a double logarithmic 
representation is used in order to identify a possible power law 
behavior such as the Pareto distribution which corresponds to 
the special form $C(w_m)=1$ for $w_m>w_0$ and $C(w_m)=(w_0/w_m)^{-\alpha}$ 
with two parameters $w_0>0$ and (typically) $\alpha>1$. In our work, 
we simply use the notation 
{\em Pareto distribution (or curve)}  for the CDF $C(w_m)$, 
for other more general cases even if there is no 
simple power law. 

However, in contrast to the Lorenz curve the behavior of the Pareto curve 
is not scale independent and in order to compare real data with 
possible RJE curves, we need to fix the global energy scale either 
by the maximal or the average value of $w_m$. Here, we show 
$C(w_m)$ as a function of $w_m/\langle w_m\rangle$ where for the 
RJS/RJE models, we simply have $\langle w_m\rangle=\eps$ (since $B=1$ and 
$E_m=w_m$). For real data the average value is more reliable since the 
maximal value of $w_m$ for a given data set may be subject to 
strong statistical fluctuations (note $w_{\max}=1$ for the RJS/RJE models 
since $B=1$).

\section{Results}

Here we compare the ENTH theory with real
Lorenz and Pareto curves
for the distributions of energy, electricity and carbon, or $CO_2$, emission
obtained from \cite{data1,data2,data3}.

\subsection{Lorenz and Pareto curves for energy}

\begin{figure}[h!]
  \centerline{\includegraphics[width=0.80\textwidth]{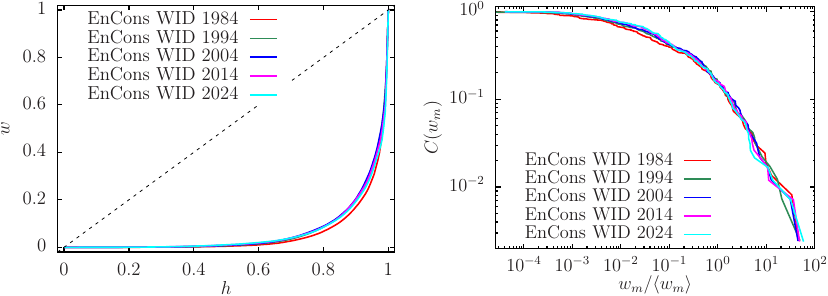}}
  \caption{{\em Left:} Lorenz curves of energy consumption of countries from WID data 
\cite{data1,data2,data3} for 5 years from 1984 to 2024. The $x$-axis corresponds to the 
cumulated fraction of households/countries ($h$) and the $y$-axis to
the cumulated fraction of wealth/energy consumption ($w$). 
The dashed black line 
corresponds to the line of perfect equipartition $w=h$. The Gini coefficients 
for the years 1984 to 2024 are $G=0.894,0.876,0.872,0.875,0.879$.
{\em Right:} Pareto curves $C(w_m)$ for the same data where 
$C(w_m)$ represents the fraction of countries with an
energy consumption larger than $w_m$ (analogous to wealth). 
The $x$-axis corresponds to rescaled 
values $w_m/\langle w_m\rangle$ where $\langle w_m\rangle = w_s$ is the 
average energy consumption for all countries which takes for the years 
1984 to 2024 the values $\langle w_m\rangle = w_s = 447,484,598,722,841$ TWh.
}
\label{fig1}
\end{figure}

In Fig.~\ref{fig1}, we show Lorenz and Pareto curves 
for the energy consumption of countries for 5 years between 
1984 and 2024 using data from the OurWorldindata website (WID) \cite{data3}. 
These curves remain very stable during the period of 40 years 
(with a slight deviation for 1984) even though 
the average energy consumption in the world has increased 
approximately by a factor $2.1$ in this period (the country average 
has only increased by a factor $1.9$ since the number of countries has 
slightly increased as well in this period). 
This is also confirmed by rather small variations of the Gini coefficient
in the range $0.894 \leq G \leq 0.972$ during 40 years 
(with the reduced interval $[0.872,0.879]$ for the last 30 years). 
The Pareto curves also remain very stable
during these 40 years. 

\begin{figure}[h!]
\centerline{\includegraphics[width=0.80\textwidth]{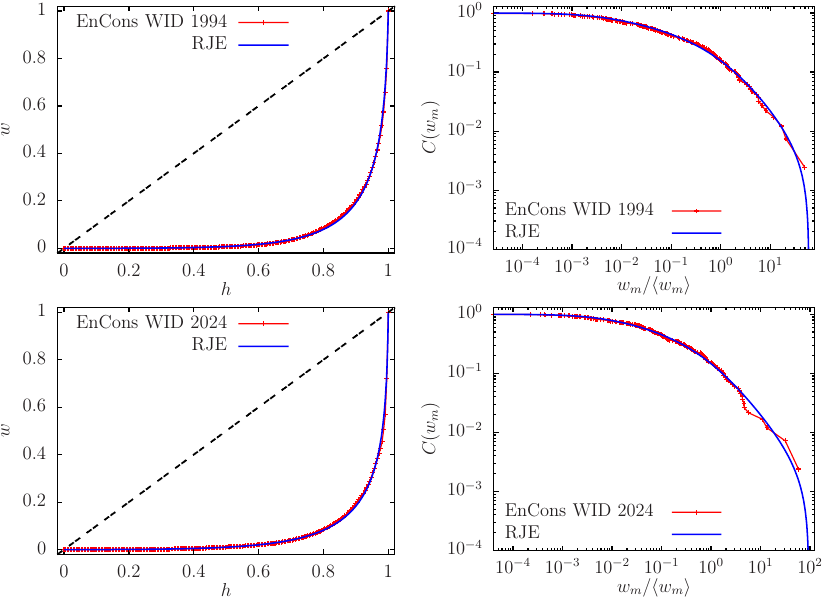}}
\caption{Lorenz curves (left panels) and 
Pareto curves (right panels) for the country energy consumption of the 
years 1994 (top) and 2024 (bottom) 
in the same style as in Fig.~\ref{fig1}. Red data points correspond to the 
WID data \cite{data1,data2,data3} for 204 (209) countries for 1994 (2024) 
and blue curves correspond to the 
theoretical RJE curves obtained by matching Gini coefficients to determine 
$\eps$ and an optimal Lorenz curve fit to determine the parameter $a$. 
The parameters for $1994$ are $G=0.875, 
\eps=\langle w_m\rangle_{\rm RJE}=0.0166, a=3.44$ 
and $\langle w_m\rangle_{1994{\rm -data}}=484$ TWh. 
The parameters for $2024$ are $G=0.879, 
\eps=\langle w_m\rangle_{\rm RJE}=0.0105, a=4.23$ 
and $\langle w_m\rangle_{2024{\rm -data}}=841$ TWh. 
}
\label{fig2}
\end{figure}

In Appendix Fig.~\ref{figA1}, we compare the Lorenz curves of 
energy consumption and production for 2022 using data 
of \cite{data1,wiki2022} which are very close to each other. 
For illustration, Fig.~\ref{figA1} also shows a Lorenz curve for the 
RJS model using the Gini coefficient for the data of 2022 energy consumption. 
This curve is different from the data but it has similar qualitative 
features. 
Therefore, in the following we
present results only for energy consumption and electricity production 
(which represents only a fraction of primary energy production) 
for which real data is more accessible to the public, especially 
in \cite{data3}. Furthermore, we  focus only on the RJE model with 
a better agreement with real data in comparison to the RJS model. 

In Fig.~\ref{fig2}, we compare 
results from the RJE model 
(with optimal parameter choice for $a$ and $\eps$) 
with real Lorenz and Pareto curves
for energy consumption of the two years 1994 and 2024 (WID data \cite{data3}) 
which are in nearly perfect agreement. We have verified that 
the RJE and real data curves for the other 
years 1984, 2004, 2014 (not shown in Fig.~\ref{fig2}) have a
similar agreement. 
In particular, we note that this agreement also holds for 
the Pareto curve on the whole range of $C(w_m)$ variations 
which cannot be reduced to a simple algebraic decay.

\begin{figure}[h!]
\centerline{\includegraphics[width=0.90\textwidth]{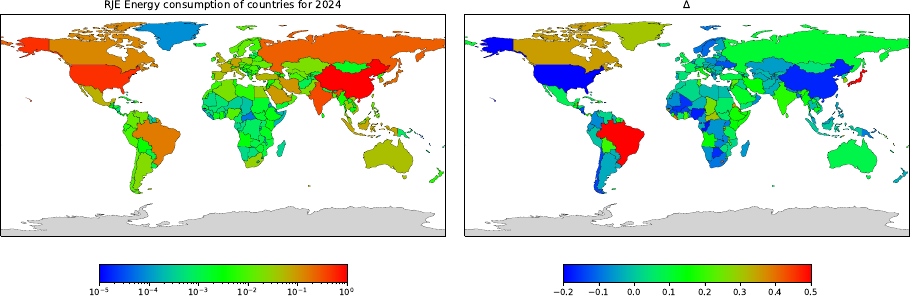}}
\caption{{\em Left:} World map of effective RJE energy consumption values 
$w_m^{\rm (RJE)}$
for 209 countries 
computed from the blue RJE Lorenz curve (in left bottom panel) 
of Fig.~\ref{fig2} at $a=4.23$, $\varepsilon =0.0105$ 
corresponding to the best RJE fit of real 2024 WID energy consumption data 
\cite{data3} 
(see text and Fig.~13 of \cite{wth2} for details). 
The color values shown in the color bar correspond to the ratio 
$w_m^{\rm (RJE)}/w_{\max}^{\rm (RJE)}$ where the maximal value 
corresponds to China. 
The blue color for the minimal value has been attributed to $10^{-5}$ 
for an optimal visibility and only a few very small countries have 
values in the interval $[10^{-6},10^{-5}]$ which are also shown as blue. 
{\em Right:} Logarithmic (or relative) difference 
$\Delta=\ln(w_m^{\rm (RJE)})-
\ln(w_m^{\rm (data)})
\approx \delta w_m/\bar w_m$ (with 
$\delta w_m=w_m^{\rm (RJE)}-w_m^{\rm (data)}$ and 
$\bar w_m=(w_m^{\rm (RJE)}+w_m^{\rm (data)})/2$)
between (sum normalized) RJE values $w_m^{\rm (RJE)}$ 
and (sum normalized) real 2024 WID energy consumption data 
values $w_m^{\rm (data)}$.
}
\label{fig3}
\end{figure}

The RJE Lorenz curves shown in Fig.~\ref{fig2}, can be used to 
recompute effective RJE wealth (energy consumption) values $w_m^{\rm (RJE)}$ 
of countries simply by taking the RJE energy $w_m^{\rm (RJE)}=E_{k(m)}$ at 
the RJE spectral index $k(m)$ 
such that $h(k(m))=(m-0.5)/N_c$ for a given country index 
$m=1,2,\ldots,N_c$ (ordered 
with increasing value of energy consumption) and where $h(k)$ is the RJE 
household dependence on the RJE spectral index $k$ 
used for the construction of the RJE Lorenz 
curve (see 
\cite{wth2} for more details on this procedure). 

In Fig.~\ref{fig3}, we show in the left panel 
a world map figure of these effective values $w_m^{\rm (RJE)}$ 
(in logarithmic color scale and using WID data of 2024) and in the 
right panel the logarithmic (or relative) difference between 
$w_m^{\rm (RJE)}$ and real data $w_m$ (with both being sum normalized to 
unity). The countries with yearly top energy consumption are China and US 
but their RJE values are 20\% below their real consumption of 2024. Otherwise, 
for most countries these differences are rather small confirming 
the validity of the ENTH approach.

\subsection{Lorenz and Pareto curves for electricity}

In Appendix Fig.~\ref{figA2}, the Lorenz and Pareto curves 
for real WID data \cite{data3} of the year 2024 
are compared with theoretical RJE curves for the case of 
{\em electricity production} with almost excellent agreement for both 
types of curves. 
Note that the (country average or global) 
electricity production represents only a fraction (about 
18\%) of the primary energy production since the latter also includes 
fossil energy sources but the found parameters of $G$, $a$ and $\eps$ 
in Fig.~\ref{figA2} are rather similar to the parameters of Fig.~\ref{fig2} 
(a bit larger/smaller values of $a$/$\eps$ with nearly same $G$ values).

\subsection{Lorenz and Pareto curves for carbon emission}

In Appendix Fig.~\ref{figA3}, we present the Lorenz and Pareto curves
for carbon emission from real WID data \cite{data3} 
for 6 years from 1974 to 2024. For each of the two curve types 
the 6 curves remain 
stable on the scale of 50 years, in particular with very similar 
Gini coefficients $G\in[0.879,0.894]$ while the country average 
CO$_2$ emission per year has practically doubled from 1974 
($8.21\times 10^7$ tons) to 2024 ($1.76\times 10^8$ tons). 

\begin{figure}[h!]
\centerline{\includegraphics[width=0.95\textwidth]{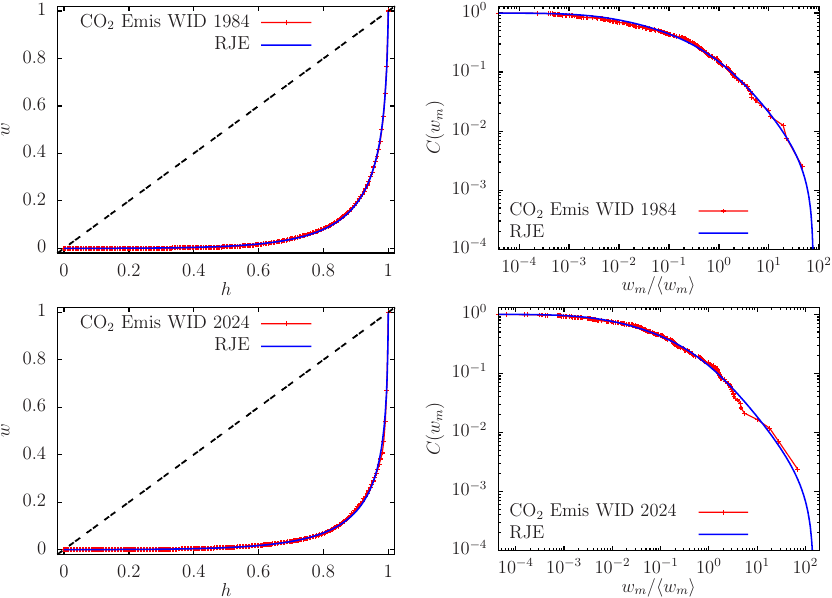}}
\caption{As Fig.~\ref{fig2} but for the cases of annual CO$_2$ emission in 
1984 (top, 200 countries) and 2024 (bottom, 213 countries) using WID data 
\cite{data1,data2,data3}. 
The RJE fit parameters for 1984 (2024) are $G=0.878$ ($G=0.893$), 
$\eps=\langle w_m\rangle_{\rm RJE}=0.0128$ 
($\eps=\langle w_m\rangle_{\rm RJE}=0.00672$) and $a=3.88$ ($a=4.76$). 
The real data country CO$_2$ emission averages are  
$\langle w_m\rangle_{1984{\rm -data}}=9.55\times 10^7$ tons  
and 
$\langle w_m\rangle_{2024{\rm -data}}=1.76\times 10^8$ tons. 
}
\label{fig4}
\end{figure}

In Fig.~\ref{fig4}, we compare in the same style as in Fig.~\ref{fig2}, 
the RJE curves to real data curves for the CO$_2$ emission of the 
two years 1984 and 2024, again with with very good agreement and 
rather similar parameter values of $G$, $a$ and $\eps$. 

Finally, we present the world map of carbon emission per country in 2024
in Appendix Fig.~\ref{figA4} in the same style of 
Fig.~\ref{fig3}, with effective RJE values $w_m^{\rm (RJE)}$ in 
the left panel and logarithmic (relative difference) of 
sum normalized values in the right panel. For most countries the 
relative difference is quite small (China has again an RJE value which 
is 20\% smaller than the real data).

\section{Discussion}

Using the ENTH description, based
on the  physical phenomenon of Rayleigh-Jeans thermalization
and condensation, we have analyzed the distribution
of total energy, electricity and carbon emission
between world countries. In several aspects
this approach is similar to the recent WTH approach of \cite{wth1,wth2}
where it was demonstrated that RJ condensation
naturally describes a huge wealth inequality in the world.
In a similar way, here we compare the results of RJ theory
with real data for energy, electricity and carbon emission of each country
over the scale of 40-50 years. Using real data \cite{data1,data2,data3}
we construct RJE Lorenz and Pareto curves 
showing that they well reproduce the real data
on a period of almost half a century. 
The advantage of Lorenz and Pareto curves is related to the fact
that they represent data in a rescaled format (if the rescaled 
wealth $w_m/\langle w_m\rangle$ is used for the Pareto curves). 
Remarkably, this rescaled representation
is very stable over 40 - 50 years
even if absolute production values of
the above quantities have significantly changed
during this time scale. We attribute this
stability in time to
the universality of the thermodynamic distribution
that can describe various systems at various temperatures. 
We point out that the RJ condensation phenomenon
naturally describes a significant inequality
in distribution of energy, electricity and
carbon emission between countries.
We demonstrate the  universal features of
the RJ thermodynamic description already used for wealth in 
\cite{wth1,wth2}. The presented results
describe the appearance of inequality 
of distributions
of various quantities in the world.
Of course, the thermodynamic description
does not imply that things cannot be changed:
it gives simply global statistical features
of a distribution over countries
or other players in a system. 
Thus e.g. carbon emission can be significantly
reduced by the world efforts
but its rescaled distribution over countries expected to remain
stable. This is not in a contradiction with the fact
that specific countries, e.g. China,
can significantly increase its carbon emission.
It is like in a thermal gas of atoms where a certain atom
due to fluctuations can change its square velocity,
but the thermal distribution of all atoms
remains the same. 
Furthermore, the RJ theory depends on the rescaled energy parameter 
$\eps$ which is essentially the ratio of average and maximal wealth (energy 
consumption, CO$_2$ emission) and the degree of inequality (Gini coefficient) 
decreases with increasing values of $\eps$ (see e.g. Fig.~4 of \cite{wth2}). 
However, for the considered data here, typically values of this parameter are 
rather stable with $\eps\sim 0.01$ and $G\approx 0.9$ for the different 
cases of energy, electricity consumption, CO$_2$ emission for different years. 

Of course, a critical mind can argue that
the fact that the  ENTH theory fits well the real Lorenz curves
may be useful but not a decisive argument
in the favor of this theory.
However, the results of this work and those in \cite{wth1,wth2}
demonstrate universal features of inequality of various
quantities in the world. We argue that
the universality of RJ thermalization and condensation
naturally describes the ubiquity of inequality in the world.
The RJ condensation gives fundamental grounds
for understanding of properties of poor and rich phases in such a variety
of systems.

We hope that the presented results
highlight in a more profound way the properties
of worldwide distribution 
of energy, electricity and carbon emission.

\section{Acknowledgments}
The authors acknowledge support from the grant
 ANR France project
NANOX $N^\circ$ ANR-17-EURE-0009 in the framework of 
the Programme Investissements d'Avenir (project MTDINA).
This work was granted access to the HPC resources of
CALMIP (Toulouse) under the allocation 2026-P0110.

\onecolumn

\clearpage

{\parindent=0cm \bf \large APPENDIX \vspace*{0.3cm}}

\setcounter{equation}{0}
\setcounter{section}{0}
\setcounter{table}{0}
\renewcommand{\theequation}{A.\arabic{equation}}
\setcounter{figure}{0}
\renewcommand{\thesection}{A.\arabic{section}}
\renewcommand\thefigure{A.\arabic{figure}}
\renewcommand\thetable{A.\arabic{table}}


Here we present additional Figures
that support the results presented in the main part of the article.


\begin{figure}[!h]
\centerline{\includegraphics[width=0.40\textwidth]{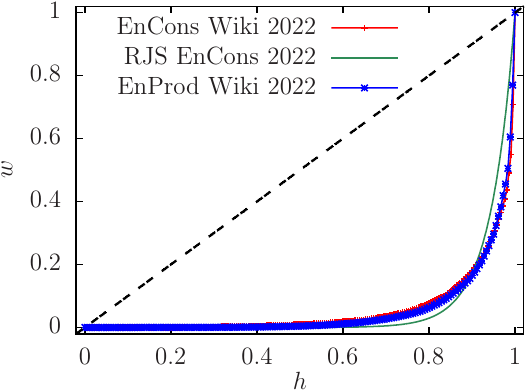}}
\caption{Lorenz curves for energy consumption (red) and production (blue)
for countries in the year 2022 taken from \cite{data1,wiki2022}; the 
Gini coefficients of both curves are $G=0.886$ (red) and $G=0.893$ (blue). 
The country average energy consumption of 2022 is 833 TWh (for 210 countries) 
and the country average energy production of 2022 is 969 TWh 
(for 181 countries) 
corresponding to a total energy consumption/production value of 
$1.75\times 10^5$ TWh. 
The green curve shows as illustration the Lorenz curve from the RJS model 
with $\varepsilon = 0.0568$ obtained from the Gini coefficient $G=0.886$ 
of the red data curve. Note that the simple RJS model corresponds to the 
limiting case $a=0$ of the RJE model. It captures the main features 
of RJ condensation but it has still visible deviations from real data 
while the RJE model with optimal parameter choice for $a$ typically 
provides very close Lorenz curves (see e.g. Fig.~\ref{fig2}).}
\label{figA1}
\end{figure}

Appendix Fig.~\ref{figA1} compares the Lorenz curves of energy production 
and consumption using data of \cite{data1,wiki2022} and it also shows 
as illustration a Lorenz curve for the 
RJS model using the Gini coefficient for the data of 2022 energy consumption. 
Note that the country average is different beween consumption and production 
which is an artificial effect of the fact that in the production data less 
countries are listed (or contribute) and the global energy and production 
values of 2022 are rather identical $\approx 1.75\times 10^5$ TWh. 

\begin{figure}[!h]
\centerline{\includegraphics[width=0.80\textwidth]{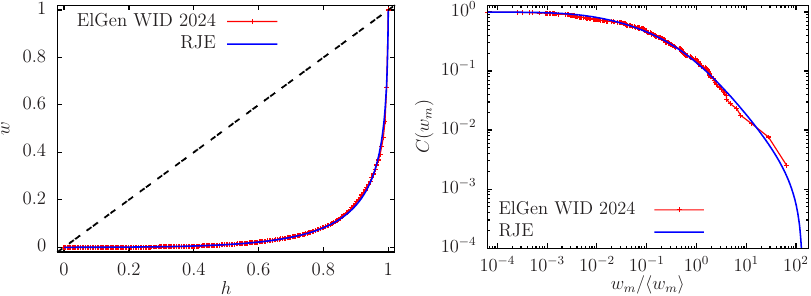}}
\caption{
As Fig.~\ref{fig2} but for the case of electricity generation in 
2024 (196 countries) using WID data \cite{data1,data2,data3}. 
The RJE fit parameters are $G=0.882$, 
$\eps=\langle w_m\rangle_{\rm RJE}=0.00733$ 
and $a=4.82$. 
The real data country electricity generation average is 
$\langle w_m\rangle_{2024{\rm -data}}=157$ TWh. 
Note that the data of Figs.~\ref{fig1} and \ref{fig2} concern 
the primary energy consumption which is not limited to electricity 
and includes also other (fossil) energy consumption with 
a larger country average of 841 TWh (for 2024). 
}
\label{figA2}
\end{figure}

Appendix Fig.~\ref{figA2} compares RJE model and real data 
for the {\em electricity production} of 2024 using WID data \cite{data3}. 

\begin{figure}[!h]
\centerline{\includegraphics[width=0.80\textwidth]{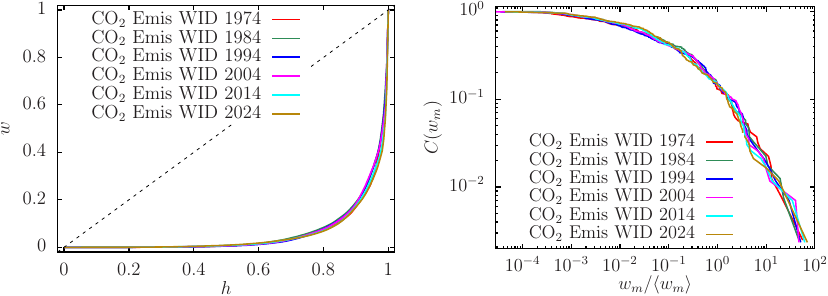}}
\caption{As Fig.~\ref{fig1} but for the country CO$_2$ emission from WID data 
\cite{data1,data2,data3} for 6 years. The Gini coefficients 
for the years 1974 to 2024 are $G=0.889,0.879,0.884,0.883,0.891,0.894$
and the country average CO$_2$ emission values for the same years are 
$\langle w_m\rangle=8.21\times10^{7},9.55\times10^{7},1.05\times10^{8},1.3\times10^{8},1.61\times10^{8},1.76\times10^{8}$ tons. 
Note that in the curves presented in 
this figure the CO$_2$ emission for international aviation
(e.g. $5.69\times10^{8}$ tons for 2024) and shipping (e.g. $6.23\times10^{8}$
tons for 2024) are not taken into account.}
\label{figA3}
\end{figure}

Appendix Fig.~\ref{figA3} presents real data Lorenz and Pareto curves 
of country CO$_2$ emission for a period of 50 years using WID data 
\cite{data3}. 

\begin{figure}[!h]
  \centerline{\includegraphics[width=0.90\textwidth]{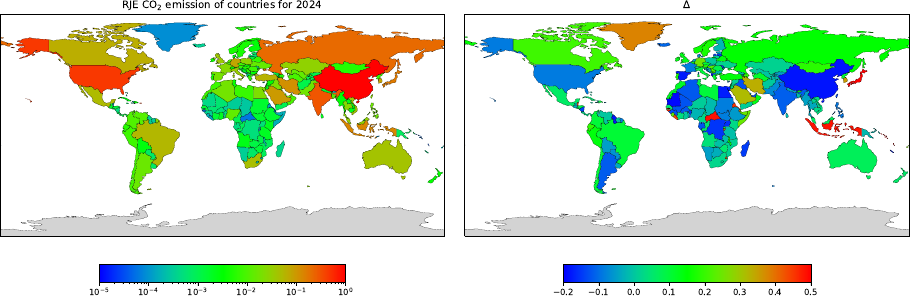}}
\caption{{\em Left:} World map of effective RJE annual CO$_2$ emission values 
$w_m^{\rm (RJE)}$ for 213 countries 
computed from the blue RJE Lorenz curve (in left bottom panel) 
of Fig.~\ref{fig4} at $a=4.76$, $\varepsilon =0.00672$ 
corresponding to the best RJE fit of real 2024 WID annual CO$_2$ emission 
data \cite{data3}. 
{\em Right:} Logarithmic (or relative) difference 
$\Delta=\ln(w_m^{\rm (RJE)})-
\ln(w_m^{\rm (data)})
\approx \delta w_m/\bar w_m$ (with 
$\delta w_m=w_m^{\rm (RJE)}-w_m^{\rm (data)}$ and 
$\bar w_m=(w_m^{\rm (RJE)}+w_m^{\rm (data)})/2$)
between (sum normalized) RJE values $w_m^{\rm (RJE)}$ 
and (sum normalized) real 2024 WID annual CO$_2$ emission 
values $w_m^{\rm (data)}$.
}
\label{figA4}
\end{figure}

Appendix Fig.~\ref{figA4} shows two panels of world map figures 
in the same style as Fig.~\ref{fig3} but for the case of 
CO$_2$ emission of 2024.

\end{document}